\PassOptionsToPackage{table,xcdraw}{xcolor}

\documentclass[conference,a4paper]{APSIPA2025}
\usepackage{amsmath}
\usepackage{graphicx}
\usepackage{subfigure}
\usepackage{multirow}
\usepackage{threeparttable}
\usepackage[backend=biber,style=ieee,]{biblatex}
\usepackage{geometry}
\usepackage{fancyhdr}

\fancypagestyle{firststyle}{
  \fancyhf{}
  \fancyhead[C]{2025 Asia Pacific Signal and Information Processing Association Annual Summit and Conference (APSIPA ASC)}
}

\begin{document}

\title{DG-SED: Domain Generalization for Sound Event Detection with Heterogeneous Training Data}




\author{
\authorblockN{Yang Xiao\authorrefmark{1}, Han Yin\authorrefmark{2}, Jisheng Bai\authorrefmark{2}, Rohan Kumar Das\authorrefmark{1}}
\authorblockA{\textit{\authorrefmark{1}Fortemedia Singapore, Singapore} \\
\textit{\authorrefmark{2}Joint Laboratory of Environmental Sound Sensing, Northwestern Polytechnical University, Xi'an, China}\\
Email: \{xiaoyang, rohankd\}@fortemedia.com, \{yinhan, baijs\}@mail.nwpu.edu.cn}
}

\maketitle
\thispagestyle{firststyle}
\pagestyle{fancy}

\begin{abstract}
This work explores domain generalization (DG) for sound event detection (SED), advancing adaptability to real-world scenarios.  Our approach employs a mean-teacher framework with domain generalization named DG-SED to integrate heterogeneous training data while preserving the SED model performance across the datasets. Specifically, we first apply mixstyle to the frequency dimension to adapt the mel-spectrograms from different domains. Next, we use the adaptive residual normalization method to generalize features across multiple domains by applying instance normalization in the frequency dimension. Lastly, we use the sound event bounding boxes method for post-processing. We evaluate the proposed approach DG-SED on the DCASE 2024 Challenge Task 4, measuring PSDS on the DESED dataset and macro-average pAUC on the MAESTRO dataset. The results indicate that the proposed DG-SED method improves both PSDS and macro-average pAUC compared to the baselines. The code will be released in due course.
\end{abstract}

\section{Introduction}

Sound event detection (SED)~\cite{sed2} involves identifying and classifying sound events from acoustic signals along with their timestamps across various environments. It is a core task in many audio-processing applications such as home surveillance and urban computing. In recent years, deep learning models~\cite{crnn,fdy,fmsg1,fmsg2,fmsg3,fmsg_dcase2024} have witnessed success in SED research. However, these models require a large amount of strongly labeled data, which is costly and time-consuming to obtain. To address these issues, weakly supervised or semi-supervised learning techniques are used and also considered under the latest edition of DCASE 2024 Challenge Task 4~\cite{dcase2024}. The current edition utilizes a new dataset namely, MAESTRO Real~\cite{maesro} that is softly labeled together with the widely used DESED~\cite{desed} dataset containing strongly labeled data for SED model development under heterogeneous training conditions. However, training a robust SED model with heterogeneous training data is challenging because the labels may not be consistent across different datasets, apart from the domain mismatch caused by differences in the multiple datasets~\cite{dcase2024}. 

Deep neural networks (DNNs) often struggle to generalize to unseen domains due to the domain mismatch mentioned above, leading to poor results in real-world applications. Domain generalization (DG)~\cite{dg,cdoa,yin2025exploring} has become an essential research topic across different fields. Techniques like feature-based augmentation, and style transfer normalization have been used to address device domain mismatch, such as in the DCASE challenge Task 1~\cite{freqmixstyle,resnorm}. Thus, employing domain generalization for heterogeneous training data from different domains for sound event detection remains underexplored. In previous editions of DCASE Challenge Task 4, researchers have explored solutions for domain mismatch. For example, domain adaptation methods have been used to efficiently exploit synthetic strong-labeled data by using domain classifiers, considering the gap between synthetic and real data in DESED~\cite{desed}. 

The DCASE 2024 Challenge Task 4 focuses on development of a single SED model for detection of sound classes in DESED and MAESTRO datasets using their training set collectively, in contrast to the previous edition in 2023~\cite{fmsg1,wild}, where separate SED model development for the two dataset was required. The goal of the SED model is still to provide event classes along with their time boundaries, even with multiple overlapping events. At the same time, this task emphasizes leveraging training data with varying annotation granularity (temporal resolution, soft/hard labels). Systems evaluated based on labels with different granularity can help to understand their behavior and robustness for various applications. In addition, the target classes in different datasets also differ, so sound labels present in one dataset might not be annotated in another. Therefore, the developed SED system needs to handle potentially missing target labels during training and perform without knowing the origin of the audio clips at evaluation time. 
\begin{figure*}[htbp]
	\centering
	\subfigure[frequency-wise statistics]{
		\begin{minipage}[b]{0.618\textwidth}
			\includegraphics[width=0.49\textwidth]{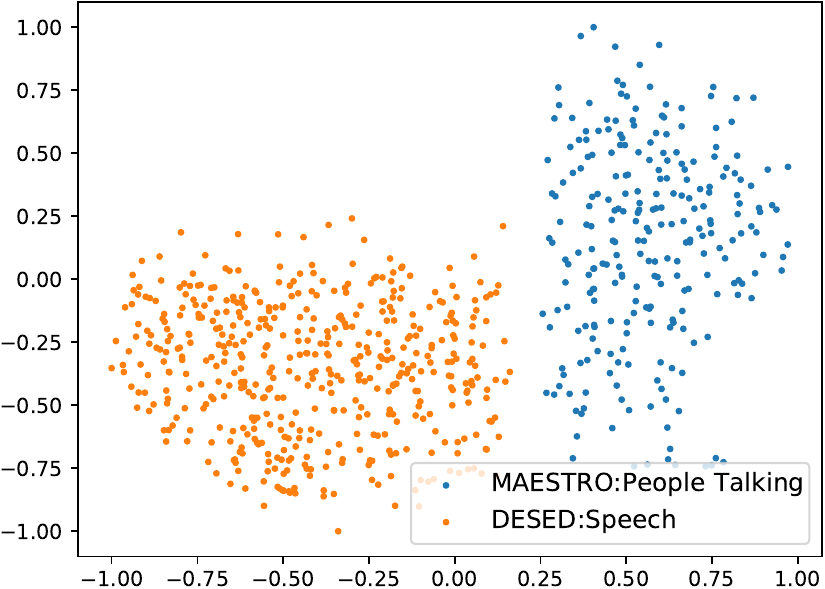} 
			\includegraphics[width=0.49\textwidth]{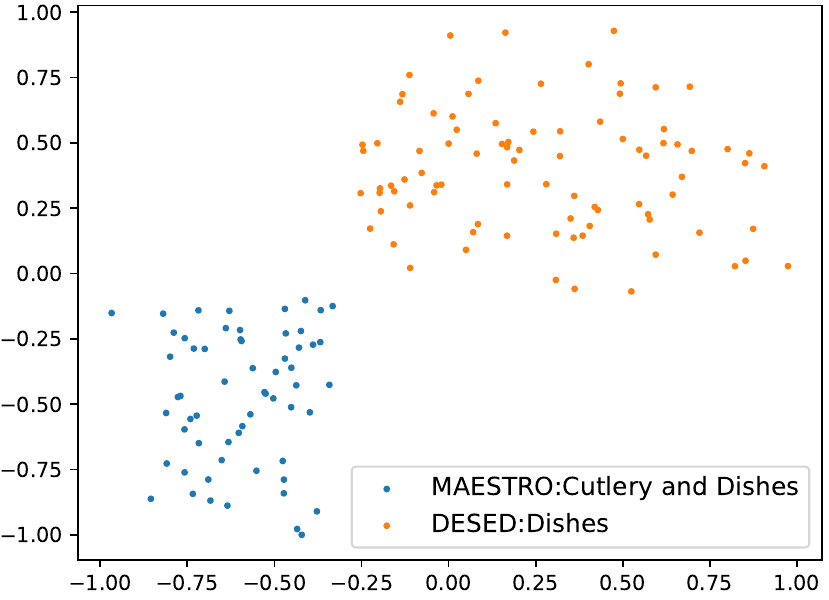}
		\end{minipage}
		\label{fig:grid_4figs_1cap_2subcap_1}
	}
    	\subfigure[channel-wise statistics]{
    		\begin{minipage}[b]{0.618\textwidth}
   		 	\includegraphics[width=0.49\textwidth]{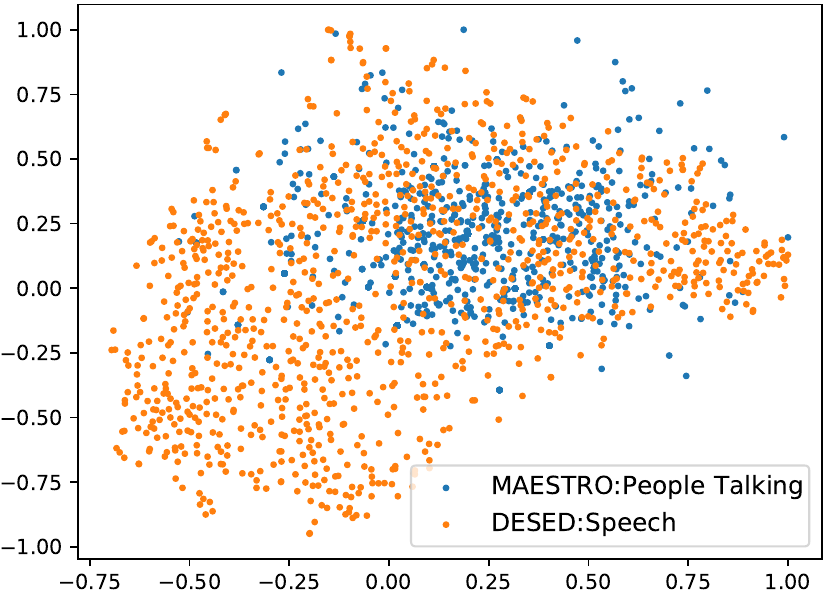}
		 	\includegraphics[width=0.49\textwidth]{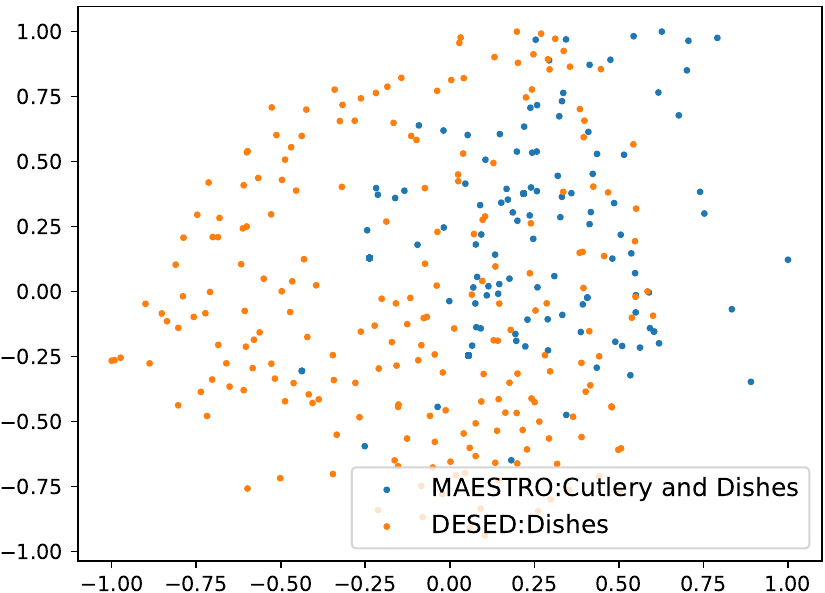}
    		\end{minipage}
		\label{fig:grid_4figs_1cap_2subcap_2}
    	}
	\caption{2D t-SNE visualizations on the DCASE 2024 Task 4 dataset using the feature map of CRNN.}
	\label{fig:tsne}
 \vspace{-6mm}
\end{figure*}
Several studies have focused on domain generalization in SED. 
The Mean Teacher method, is a key semi-supervised learning method for DG in SED~\cite{dcase2018}. However, it still struggles with distribution mismatch between synthetic and real audio~\cite{two-stage}. Domain adaptation, widely used in image and acoustic scene classification~\cite{dg-2}, offers an alternative solution.
Domain adaptation typically requires multiple stages, such as feature extraction, alignment, and mapping between source and target domains. How to achieve domain generalization in SED in a one-stage and end-to-end manner is still underexplored.

In this study, we propose a novel approach named DG-SED for SED with heterogeneous training data to address the domain mismatch issue in DCASE 2024 Challenge Task 4. We first analyze the relationship between the domain and the statistics of each feature dimension. For DG-SED, we first mix styles of training instances in the frequency dimension, resulting in novel domains being synthesized implicitly. This increases the domain diversity of the source domains and thereby enhances the generalizability of the trained model. In addition, we use the adaptive residual normalization (AdaResNorm) module to generalize features across multiple domains by applying instance normalization in the frequency dimension. Further, we use the sound event bounding boxes method for post-processing. With the above modules, the DG-SED system is expected to perform more effectively even though the training data originates from different sources. The code will be released in due course.

\section{Proposed DG-SED approach}

\subsection{Analyzing domain mismatch in heterogeneous data}
\label{analyze}
Domain mismatch in heterogeneous data is mostly due to the sources of data or how they were collected. In DCASE 2024 Task 4 Challenge, the audio files in DESED were sampled from public sources, such as Freesound and YouTube, while audio clips in MAESTRO was recorded in real-life scenarios. Although all the sound classes in DESED and MAESTRO datasets are not same, some events in DESED are mapped to similar classes in MAESTRO. For example, in DESED, ``speech" is a super-class for ``people talking," ``children's voices," and ``announcements" in MAESTRO. 
This mapping ensures the network to behave similarly during training in case of these mapped classes irrespective of their original sources. However, this class mapping may cause a domain mismatch because the audio features of these classes might differ significantly between the two datasets. 



To analyze these differences, we use a 2D convolutional neural network (CNN) as convolutional layers which are applied to extract local invariant features. We denote the input time-frequency representations of 2D CNNs as $X \in R^{F\times T}$, where $F$ and $T$ are the numbers of frequency bins and frames, respectively. With a batch size of $N$, we can represent feature maps as $M \in R^{N \times C \times F \times T}$, where $C$ is the number of channels. Following~\cite{freqmixstyle}, we utilize instance statistics across a specific dimension, i.e., mean and standard deviation (std), to analyze audio characteristics in 2D CNNs. Specifically, the frequency-wise statistics can be formulated as:
\begin{equation}
    s^{(F)} = \textrm{Concat}(\mu (F),\sigma (F))
\end{equation}
where $\mu (F)\in R^F$ and $\sigma (F)\in R^F$ are mean and std computed across $F$-axis, respectively. `Concat' stands for concatenating the
two vectors. 
Similarly, the channel-wise statistics is obtained as:
\begin{equation}
    s^{(C)} = \textrm{Concat}(\mu (C),\sigma (C))
\end{equation}
where $\mu (C)\in R^F$ and $\sigma (C)\in R^C$ are mean and std calculated across $C$-axis, respectively. 

We compare $s^{(F)}$ and $s^{(C)}$ using 2D t-SNE visualization in Figure~\ref{fig:tsne}, where the feature maps are generated by the CRNN baseline model of DCASE 2024 Task 4. It is observed that the features from different domains are better separated with frequency-wise statistics than that with channel-wise statistics, demonstrating that the frequency feature dimension carries more domain-relevant information in comparison to the channel dimension.


\subsection{Frequency-wise MixStyle}
\noindent MixStyle~\cite{mixstyle} is a common DG method motivated by the observation that the visual domain is closely related to image style. Specifically, MixStyle mixes the feature statistics of two instances with a random convex weight to simulate new styles. This helps the model generalize better across different domains by increasing the diversity of the training data. 

\begin{figure}[t]
\centering  
\includegraphics[width=0.8\columnwidth]{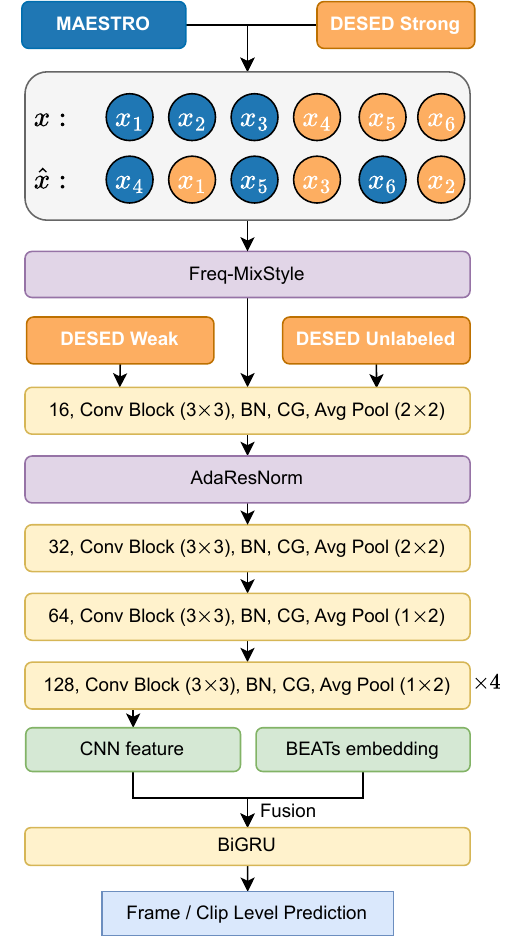}
\caption{Illustration of our proposed DG-SED method. The orange and blue labels denote the DESED and MAESTRO datasets, respectively.}
\label{fig:model}
\end{figure}

We analyzed the relationship between the domain and the statistics of each feature dimension in Section~\ref{analyze}, which revealed that the frequency feature dimension carries more domain-relevant information than the channel dimension. Therefore, we propose to use Freq-MixStyle, which normalizes the frequency bands of spectrograms and then denormalizes them with mixed frequency components from two different recordings. The mixing coefficient specifies the shape of the Beta distribution. More specifically, given an input batch \(x\), Freq-MixStyle first generates a reference batch \(\hat{x}\) from \(x\). When both datasets are given in one single batch, \(x\) is sampled from two different domains \(d\) (DESED) and \(m\) (MAESTRO), e.g., $x = [x_d, x_m]$. Then,  \(\hat{x}\) is obtained by swapping the position of \(x_d\) and \(x_m\) as shown in Figure~\ref{fig:model}, followed by a shuffling operation along the batch dimension applied to each batch. After shuffling, Freq-MixStyle computes the feature statistics, $(\mu(F, x), \sigma(F, x))$ and $(\mu(F, \hat{x}),\sigma(F, \hat{x}))$. Here, \(\mu(F, x)\) and \(\sigma(F, x)\) are the mean and standard deviation of each instance from the \(x\) across \(F\)-axis. Freq-MixStyle then generates a mixture of feature statistics by:
\begin{equation}
    \alpha_{mix} = \lambda\mu(F, x) + (1-\lambda)\mu(F, \hat{x}),
\end{equation}
\begin{equation}
    \beta_{mix} = \lambda\sigma(F, x) + (1-\lambda)\sigma(F, \hat{x}),
\end{equation}
where \(\lambda\) is an instance-specific, random weight sampled from the Beta distribution. Finally, the mixture of feature statistics is applied to the style-normalized \(x\) as:
\begin{equation}
    \text{Freq-MixStyle}(x, \hat{x})=\alpha_{mix}\odot\frac{x - \mu(F, x)}{\sigma(F, x)} + \beta_{mix}
\end{equation}

Freq-MixStyle for domain generalization is beneficial as it helps the SED model learn more robust frequency-wise features by simulating new domains during the training process. 


\subsection{Adaptive residual normalization}
While Freq-MixStyle mixes feature statistics of two instances to enhance domain generalization, we also employ adaptive residual normalization (AdaResNorm) to focus on adjusting the normalization process. The AdaResNorm is based on the frequency-instance normalization (FreqIN) which generalizes the features on multiple domains by applying instance normalization in the frequency dimension.
\begin{equation}
\text{FreqIN(x)} = \frac{x - \mu(F, x)}{\sqrt{\sigma^2(F, x) + \epsilon}}
\end{equation}
where \(\epsilon\) is an extremely small constant for numerical stability. Moreover, residual normalization adds an identity path to FreqIN with a hyper-parameter for compensating the information loss. Inspired by residual normalization, we introduce adaptive residual normalization, as shown in:
\begin{equation}
\text{AdaResNorm}(x) = (a\cdot x + (1 - a) \cdot \text{FreqIN}(x)) \cdot b + c
\end{equation}
Here \(a\), \(b\), and \(c\) are trainable parameters for balancing, scaling, and shifting, respectively. By adding trainable parameters to control the trade-off between identity and FreqIN, the normalization behavior can be adaptively adjusted by the characteristics of the data and the requirements of the domain. This can help mitigate the information loss that might occur during the MixStyle process. Additionally, AdaResNorm can enhance the robustness of the model when dealing with the diverse and mixed domains created by Freq-MixStyle. Thus, integrating AdaResNorm with Freq-MixStyle provides a more effective domain generalization by leveraging the strengths of both methods. It is noted that the adaptive residual normalization is inserted after the first convolution layer.

\subsection{Sound event bounding box-based post-processing}
Inspired by bounding box predictions in image object detection~\cite{yolo}, the SEBBs were proposed in~\cite{sebbs}, which are one-dimensional bounding boxes defined by event onset time, event offset time, sound class, and confidence. They represent sound event candidates with a confidence score. The final SED is derived by applying class-wise event-level thresholding to SEBBs' confidences. For high sensitivity or recall (few missed hits), a low detection threshold is used to detect events even when the confidence of the system is low. For high precision (few false alarms), a higher threshold detects only events with high confidence. SEBBs allow controlling system sensitivity without affecting the detection of an event's onset and offset times, unlike frame-level thresholding. 

To address the mismatch between synthetic validation and real-world test datasets, we split the 3,470 clips of the strongly annotated AudioSet to get an additional real validation dataset (373 clips). We then adopt a change-detection-based approach for SEBBs (cSEBBs) tuned on this real validation dataset. This method calculates ``delta" scores by filtering the signal and identifies peaks and troughs as tentative onsets and offsets. We merge gaps caused by minor signal variations by comparing scores with a predefined threshold. After tuning filter length and thresholds on the real validation set, cSEBBs are used as the post-processing in our system.

\begin{table*}[t]
\centering
\caption{Performance in PSDS and mpAUC of different single-systems on the DESED development set (Dev-PSDS), DESED public evaluation set (PubEval-PSDS), and MAESTRO validation set (mpAUC) including DG-SED method, and adaptive residual normalization (AdaResNorm). The joint score is based on the sum of PubEval-PSDS (cSEBBs) and mpAUC. (x) means the improvements compared without the DG-SED method.}
\label{tab:single}
\renewcommand\arraystretch{1.3}
\resizebox{\linewidth}{!}{%
\begin{tabular}{ccccccc}
\hline
\multicolumn{1}{c}{\textbf{System}} &
  \textbf{Dev-PSDS $\uparrow$} &
  \textbf{PubEval-PSDS (raw) $\uparrow$} &
  \textbf{PubEval-PSDS (cSEBBs) $\uparrow$} &
  \textbf{mpAUC $\uparrow$} &
  \textbf{Joint score $\uparrow$} &
  \textbf{Parameters}
  \\ \hline
CRNN (DCASE2024 baseline~\cite{dcase2024})  & 0.493 & 0.549 & - & 0.721 & 1.270 (-) & 1.8M\\ 
FDY-CRNN~\cite{fdy} & 0.508 &  0.596  & 0.601 & 0.728 & 1.329 (-) & 3.4M\\ \hline
CRNN + DG-SED (w/o AdaResNorm)  & 0.516 & 0.573 & 0.583 & 0.724 & 1.307 ($\uparrow$ 0.037) & 1.8M\\ 
FDY-CRNN +  DG-SED (w/o AdaResNorm) & 0.520 & 0.596 & 0.603 & 0.737 & 1.340 ($\uparrow$ 0.011) & 3.4M\\ 
\rowcolor[HTML]{C4D5EB} CRNN + DG-SED  & 0.520 & 0.574 &  0.588  & 0.726 & 1.314 ($\uparrow$ 0.044) & 1.8M\\ 
\rowcolor[HTML]{C4D5EB} FDY-CRNN +  DG-SED  & 0.526 &  0.598 & 0.604 &  0.739 & 1.343 ($\uparrow$ 0.014) & 3.4M\\ \hline
\end{tabular}%
}
\vspace{-6mm}
\end{table*}
\section{Experiment Setting}
\subsection{Implementation details}
Our implementation is based on a CRNN~\cite{crnn} from previous DCASE Task 4 challenge editions~\cite{fmsg1,dcase2022}. It is enhanced with self-supervised features from the pre-trained BEATs~\cite{beats} model. We begin by shuffling and mixing the MAESTRO and DESED strong data using Freq-MixStyle. Freq-MixStyle is used with a probability of 0.5, and a beta distribution coefficient of 0.6 is set for all experiments. After passing through the first 2D CNN layer, AdaResNorm is applied. The CRNN model includes a 2D CNN encoder with 7 convolutional layers, followed by a bi-directional GRU (biGRU) layer. Then, BEATs features are concatenated with the CNN-extracted
features before the biGRU layer. Average pooling is applied to the BEATs features to match the sequence length of the CNN encoder. Attention pooling is used to derive both clip-wise and frame-wise posteriors, and the BEATs model remains frozen during training. A mean-teacher~\cite{meanteacher,meanteacherbaseline1}  framework is utilized to leverage unlabeled and weakly labeled data, with masked softmax applied for unlabeled classes during attention pooling. We further enhance DESED dataset performance using cSEBBs after frame-level predictions. We also employed FDY-CRNN from~\cite{fdy} to demonstrate the cross-backbone ability of our domain generalization method, which uses frequency adaptive kernels to enforce frequency dependency in 2D convolutions. Compared with the baseline CRNN architecture, we replaced the standard 2D CNN with FDY-convolutional blocks.

For audio preprocessing, clips are resampled to 16 kHz mono, segmented with a window size of 2048 samples and a hop length of 256 samples, and converted into log-mel spectrograms. Clips shorter than 10 seconds are padded with silence. The training uses a batch size of 60, with specific data distribution: 1/5 MAESTRO, 1/10 synthetic, 1/10 synthetic + strong, 1/5 weak, and 2/5 unlabeled. We conduct 50 warmup epochs within a total of 300 epochs, with a learning rate of 0.001 and exponential warmup in the first 50 epochs. 

\subsection{Datasets}
\label{sec:dataset}
The DCASE 2024 Challenge Task 4 comprises of two datasets for training a single model for SED. 

{\bf The DESED dataset~\cite{desed}} comprises 10-second audio clips from domestic environments, including both real recordings and synthetic data designed to mimic these settings. The synthetic clips are generated using Scaper~\cite{scaper}, while real-world recordings are sourced from AudioSet~\cite{audioset}. The dataset includes a mix of weakly annotated (1,578 clips), unlabeled (14,412 clips), and strongly annotated data (3,470 clips), providing a diverse range of sound events for detection tasks. 

{\bf The MAESTRO Real dataset~\cite{maesro}} features a development set (6,426 clips) and an evaluation set containing long-form real-world recordings. It includes multiple temporally strong annotated events with soft labels, generated through a combination of crowdsourcing and a sliding window approach~\cite{mace}. The dataset draws recordings from the TUT Acoustic Scenes 2016 dataset~\cite{tut}, covering various real-life scenarios.

\begin{table}[t]
\centering
\caption{Performance comparison of our system with others on the DCASE 2024 Task 4 public evaluation set (PSDS) and development set (mpAUC).}
\label{tab:all}
\renewcommand\arraystretch{1.5}
\resizebox{\columnwidth}{!}{%
\begin{tabular}{ccccc}
\hline
\textbf{System}                            & \textbf{PSDS $\uparrow$}  & \textbf{mpAUC $\uparrow$} & \textbf{Joint score $\uparrow$} & \textbf{Params} \\ \hline
Chen\_NCUT\_task4\_3~\cite{Chen2024a}              & 0.549 & 0.697 & 1.246       & 17M        \\ 
Kim\_GIST-HanwhaVision\_task4\_1~\cite{Kim2024} & 0.610 & 0.686 & 1.296       & 4M         \\ 
Zhang\_BUPT\_task4\_1~\cite{Zhang2024}             & 0.572 & 0.763 & 1.335       & 10M        \\ 
\rowcolor[HTML]{C4D5EB} CRNN + DG-SED  &  0.588  & 0.726 & 1.314 & 1.8M \\
\rowcolor[HTML]{C4D5EB} FDY-CRNN + DG-SED  & 0.604 & 0.739 & 1.343       & 3.4M       \\ \hline
\end{tabular}%
}
\end{table}

\subsection{Metrics}
The DCASE 2024 Task 4 considers two metrics for evaluation. PSDS~\cite{psds,tpsds} is computed based on event onset and offset times, which are ``PSDS1" in the previous DCASE challenge and only available for DESED data, and thereby only on that fraction of the evaluation set. For MAESTRO, segment-based labels (one second) are provided, and the segment-based mean (macro-averaged) partial area under the ROC curve (mpAUC) is used as the primary metric, with a maximum FP-rate of 0.1. mpAUC is computed with respect to hard labels (threshold = 0.5) for the 11 classes listed. It is noted that the DESED and MAESTRO clips are anonymized and shuffled in the evaluation set to prevent manual domain identification.

\section{Result and Analysis}

We are first interested in evaluating the impact of the proposed DG-SED method used on the DESED dataset in terms of the PSDS metric that measures how well the system detects sound events. From Table~\ref{tab:single}, we can observe that the baseline CRNN system has a PSDS of 0.549 on the public evaluation set. And the FDY-CRNN~\cite{fdy} improves the performance with more parameters. DG-SED (without AdaResNorm) further increases the performance by improving data diversity for both CRNN and FDY-CRNN backbones. We obtain a better result with cSEBBs post-processing. Further, applying the whole DG-SED method helps with normalization to improve the PSDS. The best PSDS of 0.604 on the public evaluation set is obtained when all the modules are used, and benefits from each module. 


Then we analyze the results in terms of mpAUC, which measures performance on the MAESTRO dataset as shown in Table~\ref{tab:single}. Similar to the trend we observed in case of the DESED dataset, each method shows its contribution to enhancing the baseline performance on the MAESTRO dataset. When introducing Freq-MixStyle, and AdaResNorm together as DG-SED to FDY-CRNN backbone, we obtain the best mpAUC of 0.739 on the MAESTRO Real validation set. This confirms our proposed DG-SED method works effectively on both DESED and MAESTRO datasets. 


Finally, we compute the joint score that combines PubEval-PSDS and mpAUC scores for each system as DCASE 2024 Challenge Task 4 considers the joint score for benchmarking the systems. From Table~\ref{tab:single}, we observe that the joint score significantly improves from 1.270 to 1.314 upon introducing DG-SED with cSEBBs to the baseline CRNN system, showcasing their contributions. Also with the FDY-CRNN backbone, we obtain the best joint score, it comes at the cost of the increase in model parameters around twice that in the baseline. On the contrary, DG-SED is useful for developing low-complexity systems as it does not affect the model parameters. 

This Table~\ref{tab:all} shows that our DG-SED system, which incorporates domain generalization methods (e.g., MixStyle and Adaptive Residual Normalization) into the FDY-CRNN with cSEBBs, achieves a joint score of 1.343 on the DCASE 2024 Task 4 public evaluation set and . This score surpasses other top submissions under similar resource constraints.  We avoid direct comparisons with models with very different parameter sizes, to ensure a fair evaluation. Our model requires only 3.4 million parameters, demonstrating that it enhances accuracy while maintaining a compact structure.

\section{Conclusion}
This work demonstrates the effectiveness of DG-based techniques for SED systems trained with heterogeneous data. By integrating Freq-MixStyle, adaptive residual normalization, and cSEBBs within a mean-teacher framework, our DG-SED approach successfully adapts mel-spectrograms and generalizes features across multiple domains. Studies on the DCASE 2024 Challenge Task 4 dataset show that our DG-SED method significantly improves both the polyphonic SED score on the DESED dataset and the macro-average pAUC on the MAESTRO Real dataset, outperforming the challenge baseline. These results highlight the potential of DG-SED methods in advancing SED adaptability to real-world scenarios.

\printbibliography

\end{document}